\documentclass[AMA,STIX1COL]{WileyNJD-v2} 
\usepackage{makecell}
\usepackage{subcaption}
\articletype{}%

\received{Date Month Year}
\revised{Date Month Year}
\accepted{Date Month Year}
\copyyear{2023}
\startpage{1}

\begin{document}

\title{Subgroup identification using individual participant data from multiple trials on low back pain}

\author[1]{Cynthia Huber}

\author[1]{Tim Friede}

\authormark{Huber and Friede}
\titlemark{Subgroup identification in low back pain data}

\address[1]{\orgdiv{Department of Medical Statistics}, \orgname{University Medical Center Göttingen}, \orgaddress{Göttingen, \country{Germany}}}

\corres{Corresponding author Cynthia Huber \email{cynthia.huber@med.uni-goettingen.de}}


\abstract[Abstract]{Model-based recursive partitioning (MOB) and its extension, metaMOB, are potent tools for identifying subgroups with differential treatment effects. 
In the metaMOB approach random effects are used to model heterogeneity of the treatment effects when pooling data from various trials. 
In situations where interventions offer only small overall benefits and require extensive, costly trials with a large participant enrollment, leveraging individual-participant data (IPD) from multiple trials can help identify individuals who are most likely to benefit from the intervention.
We explore the application of MOB and metaMOB in the context of non specific low back pain treatment, using synthesized data based on a subset of the individual participant data meta-analysis by Patel et al \cite{Patel}. 
Our study underscores the need to explore heterogeneity in intercepts and treatment effects to identify subgroups with differential treatment effects in IPD meta-analyses. }

\keywords{subgroup identification, IPD, meta-analysis, heterogeneity, model-based recursive partitioning, metaMOB, GLMM-tree}


\maketitle

\renewcommand\thefootnote{}


\section{Introduction}\label{intro}
For interventions with small to moderate treatment benefits, investigators often aim to identify particular patient subgroups that could potentially derive a greater treatment advantage\cite{ema_subgroup}. If in addition to the smaller overall treatment benefit the trial sizes are small due to time and cost reasons, pooling data from several trials for the analyses is attractive. \\
For non-specific low back pain (NSLBP) a repository of individual participant data from 19 completed NSLBP trials testing similar nonpharmacotlogical interventions was set up to investigate which patients are more likely to benefit from treatment in terms of the back-related disability.\\
For subgroup identification, Seibold et al \cite{mobS} investigated the use of model-based recursive partitioning (MOB), a versatile tree-based method combining parametric models with recursive partitioning \cite{mobZ}. MOB showed an overall good performance in neutral comparison studies \cite{Huber.2019, sies, alemayehu, loh_comparison} assessing the performance of subgroup identification approaches in precision medicine. For the identification of subgroups based on individual participant data from multiple studies metaMOB, a generalised linear mixed model tree approach \cite{Fokkema2017}, was investigated by Huber et al \cite{Huber.2022}. Analyzing IPD in meta-analysis, including the use of tree-based methods like metaMOB accounting for two distinct types of heterogeneity, baseline and treatment effects, is essential in the context of IPD meta-analysis \cite{Huber.2022}.


In this article, we illustrate the use of different MOB approaches that use different approaches for accounting for the different types of heterogeneity on synthesized low back pain data based on the data set collected by Patel et al \cite{Patel}.

\section{Low Backpain Data}\label{low-backpain}

Patel et al \cite{Patel} collected data from 19 completed NSLBP trials to identify participant characteristics predicting clinical response to treatments for low back pain.
For this article, we synthesised a subset of the repository. We synthesized data according to
NSLBP data from 1780 synthesised individuals of four two-arm trials included in the repository. For synthesizing the data we used the R-package \texttt{synthpop}.\\
Table \ref{tab:tab1} presents demographics and the Roland Morris
Disability Questionnaire (RMDQ) stratified by synthesized trial. The RMDQ
score ranges from 0 to 24 with higher values indicating a poorer outcome
associated with back pain. The table illustrates that the trials are of
different sizes.  The largest trial, trial with ID 1, includes over
\(1000\) patients whereas trial 3 involves only \(53\) patients.
RMDQ was measured at different follow-up visits. Our analyses are based on the data summarized in Table \ref{tab:tab1}. The synthesized analysis data includes four studies in
which RMDQ is measured at baseline and at least at one follow-up visit.
As outcome we considered the last observation of RMDQ of each patient.

\begin{table}[h]
\caption{Summary of the synthesized analysis set}
\label{tab:tab1}
\centering
\begin{tabular}{|l|l|l|l|l|l|}
\hline
\textbf{Variable} & \textbf{N} & \textbf{Trial 1, N = 1,087} & \textbf{Trial 2, N = 232} & \textbf{Trial 3, N = 53} & \textbf{Trial 4, N = 176} \\
\hline
\textbf{Male} & 1,548 & 473 (44\%) & 131 (56\%) & 37 (70\%) & 61 (35\%) \\
\textbf{Intervention }& 1,548  & 805 (74\%) & 110 (47\%) & 22 (42\%) & 82 (47\%) \\
\textbf{RMDQ\_0} & 1,537 & 8.0 (5.0, 12.0) & 13.0 (10.0, 17.0) & 14.0 (9.0, 16.0) & 5.0 (4.0, 8.0) \\
\textbf{RMDQ at last follow-up} & 1,523 & 4.0 (1.0, 8.0) & 1.0 (0.0, 3.0) & 5.0 (3.0, 8.0) & 2.0 (0.0, 5.0) \\
\textbf{Age in years} & 1,548 & 44 (35, 52) & 41 (32, 50) & 44 (34, 53) & 40 (34, 49) \\
\hline
\end{tabular}
\end{table}

Table \ref{tab:tab1} illustrates that the baseline characteristics, sex, age and RMDQ at baseline slightly differ between trials. A linear model not accounting for any heterogeneity by, e.g. adjusting for trial, does not reveal a significant benefit of the intervention with regard to the RMDQ reduction from baseline. However, adjusting for trial in a linear model and therefore accounting for heterogeneity in the baseline outcome between trials, shows a significant reduction of the RMDQ in the intervention (see Table \ref{tab:lm}). 
Furthermore, we applied linear mixed models to the entire pooled population, incorporating the treatment as a random effect, and incorporating both the treatment and trial indicator as random effects. This analysis aimed to explore the heterogeneity between trials.
We found evidence for heterogeneity in the baseline. The variance of the random treatment effect incorporated in the linear mixed model with random trial and treatment effect (result not shown here) is estimated to be zero. However, the model incorporating the trial indicator as fixed effect and the treatment as random effect only results in a variance of the random treatment effect larger than zero, see Table \ref{tab:lm} (c). 

\begin{table}[h]
\caption{Treatment effect estimated by (a) linear model and (b)linear model including the trial indicator as fixed effect}
\label{tab:lm}
\begin{minipage}{0.5\textwidth}
\centering
\begin{subtable}{\textwidth}
\caption{Unadjusted linear model}
\begin{tabular}{|l|l|l|l|l|}
\hline

\textbf{Characteristic} & \textbf{Coefficient} & \textbf{95\% CI} & \textbf{p-value} \\
\hline
Intervention & -0.17 & -0.69, 0.35 & 0.5 \\
\hline
Baseline RMDQ & 0.26 & 0.21, 0.31 & $<0.001$ \\
\hline
\end{tabular}
\end{subtable}
\end{minipage}%
\begin{minipage}{0.5\textwidth}
\centering
\begin{subtable}{\textwidth}
\caption{Linear model adjusted for trial}
\begin{tabular}{|l|l|l|l|l|}
    \hline
    \textbf{Characteristic} & \textbf{Coefficient} & \textbf{95\% CI} & \textbf{p-value} \\
    \hline
     \hline
Intervention & -1.1 & -1.6, -0.64 & $<0.001$ \\
Baseline RMDQ & 0.40 & 0.35, 0.46 & $<0.001$ \\
Trial indicator  &  &  & \\
2 & -5.8 & -6.5, -5.1 & $<0.001$ \\
3& -2.2 & -3.6, -0.87 & $0.001$ \\
4 & -1.3 & -2.1, -0.56 & $<0.001$ \\
\hline

  \end{tabular}
\end{subtable}
\end{minipage}

\begin{minipage}{0.5\textwidth}
\centering
\begin{subtable}{\textwidth}
\caption{Linear mixed model with treatment as random and trial indicator as fixed effect}
\begin{tabular}{|l|l|l|l|}
\hline
\textbf{Characteristic} & \textbf{Coefficient} & \textbf{95\% CI} & \textbf{p-value} \\
\hline
Intervention & -1.1 & -3.0, 0.82 & 0.12 \\
Baseline RMDQ & 0.40 & 0.35, 0.46 & $<0.001$ \\
Trial indicator  &  &  & \\
2 & -6.4 & -13, -0.22 & 0.048 \\
3 & -2.4 & -7.3, 2.4 & 0.15 \\
4 & -1.5 & -9.3, 6.3 & 0.3 \\

\hline
Variance for intervention & 0.137 &  &  \\
\hline
\end{tabular}
\end{subtable}
\end{minipage}
\end{table}

\section{Subgroup identification}\label{subgroup-identification}
We illustrate the identification of subgroups in NSLBP with differential treatment effects using MOB.
MOB and its extensions, as metaMOB or palmtree \cite{palmtree}, are based on generalised linear models. To describe the different variations of MOB and the underlying models we assume that that the outcome RMDQ is denoted by $y$. Furthermore, the treatment indicator is denoted by $t$, the covariates age, sex and RMDQ at baseline are denoted by $x_{age}$, $x_{sex}$ and $x_{RMDQ_0}$. The baseline covariates $x_{age}$, $x_{sex}$ and $x_{RMDQ_0}$ are considered as potential splitting variables and are therefore not involved in the underlying regression model. The analysis includes four trials $k=1,\ldots, 4$. The corresponding trial indicator is denoted by $x_{trial}$.\\
For MOB the outcome of each subgroup $j$ and trial $k$ is modelled by:
\begin{equation}
\label{eq:mob}
y_{jk}=\gamma_{j}+\beta_j t.
\end{equation} 
MOB does not account for the data being pooled from different trials; therefore, the right hand side of the equation does not depend on $k$.
Adjusting for the different trials by fixed effects using MOB is referred to as MOB-SI. SI refers to a \textbf{s}tratified \textbf{i}ntercept as for each trial a separate intercept is estimated. The linear model for RMDQ fitted in each subgroup $j$ based on the method MOB-SI is: 
\begin{equation}
\label{eq:mobsi}
y_{jk}=\gamma_{jk}+\beta_j t,
\end{equation} 
with $\gamma_{jk}$ describing the subgroup and trial-specific fixed intercept. For both MOB and MOB-SI the \texttt{lmtree} function of the \texttt{partykit} package is used.
Addressing heterogeneity in the baseline with \textbf{r}andom \textbf{i}ntercepts can be achieved by applying MOB-RI. MOB-RI is based on GLMM-trees. Therefore, for analysing the NSLBP data with RMDQ as outcome a linear mixed model is fitted to each subgroup $j$ and trial $k$:
\begin{equation}
\label{eq:mobri}
y_{jk}= \gamma_{j}+\beta_jt_k+b_{0k} \text{ with } b_{0k}\sim \mathcal{N}(0,\tau_0^2).  
\end{equation}

The random intercept $b_{0k}$ is considered to be the same for each subgroup.
It is possible to additionally account for heterogeneity in the treatment effect by using the following metaMOB approaches:\\
metaMOB-RI

\begin{equation}
\label{eq:metaMOBri}
y_{jk}= \gamma_j+\beta_jt_k+b_{0k}+b_{1k} \text{ with } b_{0k}\sim \mathcal{N}(0,\tau_0^2)\text{ and } b_{1k}\sim \mathcal{N}(0,\tau_1^2),  
\end{equation}
and metaMOB-SI:
\begin{equation}
\label{eq:metaMOBsi}
y_{jk}= \gamma_{jk}+\beta_jt_k+b_{1k} \text{ with }  b_{1k}\sim \mathcal{N}(0,\tau_1^2).
\end{equation}
The approaches involving random effects, MOB-RI, metaMOB-SI and metaMOB-RI are fitted using the \texttt{lmertree} function of the \texttt{glmertree} package.\\
An alternative to both the MOB-RI and MOB-SI approaches, which address heterogeneity in baseline by using subgroup and trial-specific intercepts (MOB-SI) or random intercepts for the trial indicator (MOB-RI), is the Generalized Linear Model Trees with global additive effects, referred to as \textit{palmtree} \cite{palmtree}. 
In contrast to MOB-RI which assumes the random intercepts to be constant across subgroups, \textit{palmtree} includes the treatment indicator in the model similar to MOB-SI, but assumes these intercepts to be the same across the identified subgroups:
\begin{equation}
\label{eq:palmtree}
y_{jk}=\gamma_k+\beta_j t.
\end{equation} 

The regression models used for the metaMOB approach (Equation \ref{eq:metaMOBri} and \ref{eq:metaMOBsi}) are in alignment with the models typically used in random-effects meta-analysis, here the normal-normal hierarchical model.
 
\subsection{MOB}\label{mob}
\begin{figure}
\centering
\includegraphics[width=0.95\textwidth]{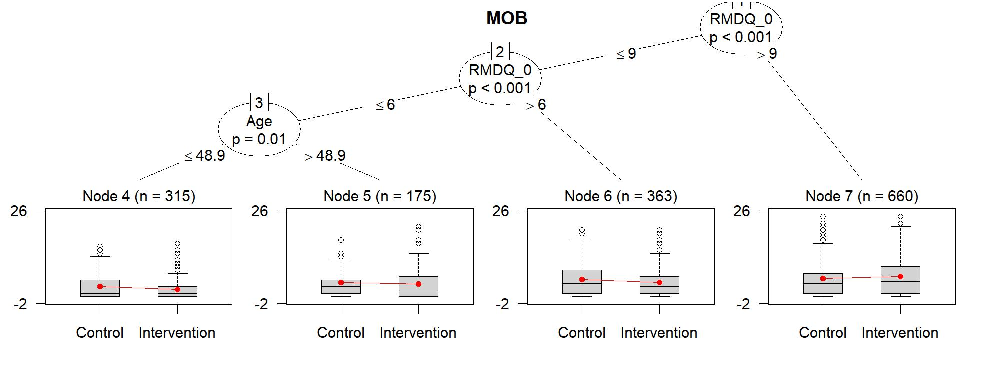}
\caption{Tree obtained by MOB }
\label{fig:mob}
\end{figure}

\begin{figure}
\centering
\includegraphics[width=0.95\textwidth]{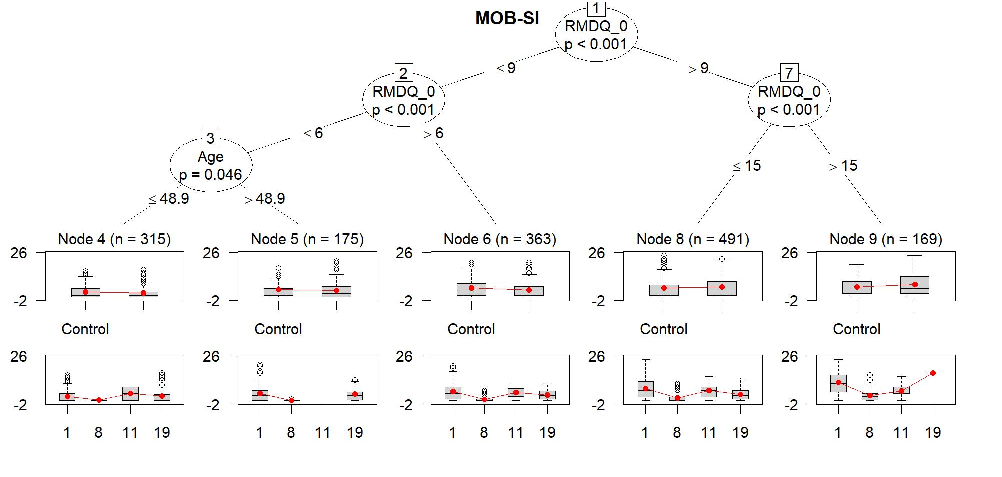}
\caption{Tree obtained by MOB-SI }
\label{fig:mobSI}
\end{figure}

We applied MOB with the default values for the different arguments of the \texttt{lmtree} function to the NSLBP data. MOB identifies four subgroups as illustrated by Figure \ref{fig:mob}. The subgroups are defined by the baseline values of RMDQ and Age. The linear model, which incorporates the treatment indicator only (Equation \ref{eq:mob}), shows a significant ($<0.05$) reduction in RMDQ for the intervention group compared to the control group within the first subgroup. In the remaining three subgroups the treatment effects were not significantly different from zero. 
However, as seen in the previous analyses (see Table \ref{tab:lm}) adjustment for the trial indicator seems appropriate. MOB-SI accounts for heterogeneity in the baseline by adjusting for the trial indicator, see Equation \ref{eq:mobsi}. The result of this approach is illustrated in Figure \ref{fig:mobSI}. In contrast to MOB, MOB-SI identifies more subgroups, namely five. MOB-SI additionally partitions the group of participants with baseline RMDQ values larger than 9 into two additional subgroups. MOB-SI estimates the treatment effect separately in each of the identified subgroups using a linear model with treatment and trial as factors, resulting in p-values for the treatment effect within two subgroups (node 4 and node 6) that are smaller than $0.05$.
Both of these subgroups estimate a reduction of the RMDQ score of the intervention compared to the control indicating a treatment benefit.

An alternative to MOB with stratified intercept for the trial is the
GLMM-tree algorithm with a random intercept to account for baseline
heterogeneity. This approach is referred to by MOB-RI \cite{Huber.2022}, which imposes a restriction on the heterogeneity of the baseline by assuming a certain distribution of its random effects, see Equation \ref{eq:mobri}. Nevertheless, the subgroups identified by MOB-RI are the same as for MOB-SI. 

\subsection{metaMOB}\label{metamob}
\begin{figure}
\centering
\includegraphics[width=0.95\textwidth]{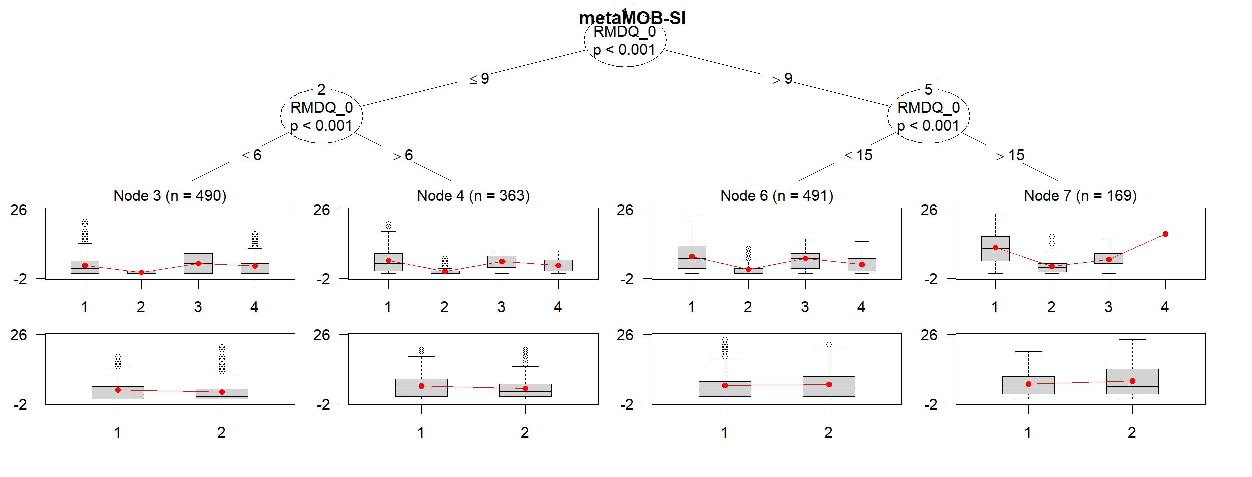}
\caption{Tree obtained by metaMOB-SI }
\label{fig:metamobSI}
\end{figure}

Heterogeneity in the treatment was not assumed by the models 
MOB, MOB-RI and MOB-SI. The analysis of the pooled population with a linear mixed model presented in Table \ref{tab:lm} suggests that heterogeneity should also be considered for the MOB models. \\
The estimated tree of metaMOB-RI is identical to the ones estimated by MOB-SI and MOB-RI (see Figure \ref{fig:mobSI}. The tree estimated by \textit{palmtree} is also consistent with those estimated by MOB-SI, MOB-RI, and metaMOB-RI.\\
When fitting the model that incorporates heterogeneity in both baseline and treatment effects by employing random effects for both, the estimated variance of the random treatment effect is equal to zero. \\
Huber et al \cite{Huber.2022} recommended to use metaMOB-SI as it is the most flexible approach and showed the best performance regarding different measures, e.g. false discovery rate across different scenarios. Applying metaMOB-SI to NSLBP identifies four subgroups, only. Therefore, it differs from the results obtained by MOB-SI, MOB-SI and metaMOB-RI. The tree obtained by metaMOB-SI is illustrated in Figure \ref{fig:metamobSI}. The estimated treatment effect in metaMOB-SI underlying linear mixed model is not significant for any of the identified subgroups. In node 4 (as denoted in the figure) the largest treatment benefit is observed. The estimated treatment effect for node 4 is $-1.44$ (p-value: $0.07$).
The variance of random treatment effect $b_1$, which is assumed to be constant over the identified subgroups, is estimated to be $0.59$ in the underlying mixed model of metaMOB-SI. 

\section{Discussion}
When identifying subgroups with differential treatment effects based on data coming from multiple trials, it is crucial to account for the heterogeneity between these trials \cite{Huber.2022, Sela2012, Mistry2018}. 
In this manuscript, we illustrated how different modelling approaches for heterogeneity in the intercepts and the treatment effects can lead to different results based on a subset of synthesized NSLBP data. Due to a limited number of baseline covariates considered to define the subgroups, the results of the different procedures only slightly differ. The approach recommended by Huber et al \cite{Huber.2022} is the only approach whose results differed from the other approaches accounting for (different types of) heterogeneity. MetaMOB-SI identified fewer subgroups compared to the other approaches which aligns with the results obtained by the simulation study in Huber et al.\cite{Huber.2022}: Approaches that make assumptions resembling the true underlying heterogeneity structure often result in less complex trees. \\
Although the metMOB-SI approach was recommended it also comes with its limitations regarding the estimability of the regression coefficients for the dummy coded trial indicator variable, the stratified intercept. The application of MOB-SI and metaMOB-SI to the NSLBP data encountered the same problem. In node 9 of MOB-SI and in node 7 of metaMOB-SI, only a small number of patients from trial 4 were included, not allowing to accurately estimate separate intercepts. 
The underrepresentation of the trial indicator, included as a fixed effect in the subgroup-specific models of MOB-SI and metaMOB-SI, hinders the potential for further splitting on these nodes because the underlying model and, consequently, the splitting criterion cannot be calculated. In such cases the assumption of random intercepts as in MOB-RI and metaMOB-RI provides an advantage. \\
Additionally, the number of trials included in this illustrative analysis is also small. Including a smaller number of trials in a meta-analysis increases the likelihood of heterogeneity estimates for between-trial treatment effects being equal to zero \cite{Friede.2017}. Variance estimated of the random treatment effect of zero were observed for the linear mixed model on the overall population as well as for the model including the identified subgroups of metaMOB-RI. For the mixed model defined in Equation \ref{eq:metaMOBsi} and therefore metaMOB-SI's underlying model the random treatment effect variance was not estimated to be equal to zero. Furthermore, some of the methods might not account appropriately for uncertainty in estimating the heterogeneity with only a few studies. A Bayesian approach with a weakly informative prior for the heterogeneity might have favourable properties.
The assumptions made on the between-trial heterogeneity therefore influence both the subgroups identified and the treatment effect estimates.

\bibliography{BIB}

\end{document}